# Langmuir wave self-focusing versus decay instability

Harvey A. Rose[a]

Los Alamos National Laboratory

Electron trapping in a finite amplitude Langmuir wave (LW) leads to a frequency shift, $\Delta\omega_{TP} < 0$, and reduced Landau damping. These may lead to modulational instability. Its growth rate and damping threshold, due to escape of trapped electrons at rate $\nu$, are calculated for the first time in the short wavelength regime. If the background plasma is in thermal equilibrium, it is shown that this trapped particle modulational instability (TPMI) is not possible when $k\lambda_D > 0.46$, while for $0.33 < k\lambda_D < 0.46$, TPMI requires that the fluctuation wavevector have a component perpendicular to **k**, the LW wavevector, with $\lambda_D$ the electron Debye length. Its nonlinear evolution leads to self-focusing. Comparison is made with a re-evaluated LW ion acoustic decay instability (LDI): compared to classical estimates, the new LDI threshold is lowered by primary LW $\Delta\omega_{TP}$ since frequency matching leads to wavenumber and hence damping reduction of the daughter LW. For parameters estimates relevant to a recent stimulated Raman scatter experiment (Kline *et al.*, submitted to PRL), the LDI and TPMI thresholds cross in the range $0.28 < k\lambda_D < 0.34$, consistent with the observed LDI regime change. However, if $\nu$ exceeds a critical value, estimated to be order 1% of the electron plasma frequency, then TPMI is not possible at any wavenumber.

---

[a] Electronic mail: har@lanl.gov

52.35.Mw, 52.35.Sb, 52.38.Hb

# I. INTRODUCTION

A Langmuir wave's angular frequency, $\omega$, depends on its amplitude, represented here by its electrostatic potential, $\phi$, through a variety of mechanisms. One example is the wave's ponderomotive force: it induces a low frequency plasma density fluctuation, $\delta n \propto \phi^2$, and hence a shift in the wave frequency since the electron plasma frequency, $\omega_p$, is, of course, density dependent. When combined with ion acoustic wave dynamics, as in Zakharov's model[1] of Langmuir wave (LW) turbulence, this frequency shift leads to the LW ion acoustic decay instability[2] (LDI) and LW collapse[3]. If the LW's wavenumber, $k$, is much greater than $k_* \equiv (2/3)(c_s/v_e)k_D$ ($c_s$ is the ion acoustic speed, $v_e$ the electron thermal speed, $\lambda_D = v_e/\omega_p$ the electron Debye length and $k_D\lambda_D = 1$), then, as is well known, the LDI growth rate[4], $\gamma_{LDI} \propto k\phi\sqrt{k}$, for small $\phi$. The ponderomotive force also induces a modulational instability whose growth rate varies as $\phi^2 k^2$, so that LDI is dominant for small enough $\phi$.

Another example is the LW trapped particle frequency shift[5][6][7], $\Delta\omega_{TP}$, which for small $\phi$ satisfies $\Delta\omega_{TP} = h(k\lambda_D)\omega_b$, $\omega_b = kv_e\sqrt{e\phi/T_e}$, the electron bounce frequency, $T_e$ the electron temperature, with $h < 0$. In one dimension (1D), for $k\lambda_D << 1$ and absent damping, it has been shown to lead[8] to the trapped particle modulational instability (TPMI), with growth rate $\gamma_{TPMI} = \left| \Delta\omega_{TP}/4 \right|$. Current interest in high temperature, low density plasma, in particular, the quantitative measurement[9] of large $k\lambda_D$ Langmuir waves generated by backward stimulated Raman scatter (SRS), motivates the main subject of the current study: extension of TPMI theory into the short wavelength regime.

After reviewing the basic model in section II.A, an analytic large-$k\lambda_D$-valid expression for $\Delta\omega_{TP}$ is reviewed in section II.B and compared with various perturbative results. Comparison with a LW's ponderomotive force induced frequency shift is presented in section II.C.

Various TPMI regimes are discussed in section III.A. Positive LW dispersion[8] is required for instability since $\partial(\Delta\omega_{TP})/\partial\phi < 0$. If the LW propagates in the "$z$" direction, then positive dispersion requires either $\partial^2\omega/\partial k_x^2 = \partial^2\omega/\partial k_y^2 > 0$, and/or $\partial^2\omega/\partial k_z^2 > 0$. If $\phi$ is small, $i.e.$, $e\phi/T_e \ll 1$, it is shown in section III.A that these inequalities are satisfied for $k\lambda_D < 0.33$. But LW dispersion is mixed for $0.33 < k\lambda_D < 0.46$: $\partial^2\omega/\partial k_x^2 > 0$ while $\partial^2\omega/\partial k_z^2 < 0$. For $k\lambda_D > 0.46$, dispersion is negative in all directions. 3D wave packet collapse is possible in the first wavenumber regime; 2D self-focusing in the $x$-$y$ plane in the second, mixed dispersion, regime; and TPMI is not possible in the third. If $\phi$ is not small, it will be shown that these regimes move to smaller $k\lambda_D$. In the second regime, 1D models would miss TPMI that might otherwise occur in 2D and 3D models.

Since LDI has been unambiguously observed[9] in the large $k\lambda_D$ regime, comparison of its growth rate (section III.B) and threshold (section III.C) with TPMI's is natural. It will be shown that the TPMI amplitude damping threshold *decreases* rapidly with increasing $k$, because $|\Delta\omega_{TP}|$ is a rapidly increasing function of $k$, while the residual LW damping, owing to the loss of trapped electrons, is not. Conversely, the *classical* LDI amplitude threshold increases rapidly with $k$, because the daughter LW's Landau damping increases rapidly with $k$, and therefore it will be smaller than the TPMI threshold for small enough $k\lambda_D$ (small enough $T_e$), and *visa versa*. However, the LDI frequency/wavenumber matching constraints are altered by the trapped particle frequency shift of the primary LW in such a way as to dramatically reduce the LDI threshold, compared to its classical, value for large $k\lambda_D$. This new LDI threshold is compared with the TPMI threshold in section III.C.4.

## II. TRAPPED PARTICLE FREQUENCY SHIFT REDUX

Analytic expressions for the trapped particle frequency shift, $\Delta\omega_{TP}$, valid for small $k\lambda_D$ and small $e\phi/T_e$, have been derived by various authors (see references 5-7). An expression also valid for finite $k\lambda_D$ and $e\phi/T_e$ was later obtained by Rose and Russell[10] (R&R). These results are now reviewed.

### A. The basic model: its resonance and damping

The electron dynamics model (see Eq. (5) of R&R) is given by the 1D Vlasov equation with external source (potential), and a term that represents coupling to a background plasma. Let the Langmuir wave source be given by $\mathrm{Re}\,\phi_0\exp[i(kz-\omega t)]$, with $\phi_0$ constant. Then in equilibrium[11], the harmonic component of the total potential envelope, $\phi$ =electrostatic + $\phi_0$, with the electrostatic component obtained from Poisson's equation, is given by

$$\phi = \phi_0/\varepsilon. \qquad (1)$$

This defines the nonlinear dielectric function, $\varepsilon$. Coupling to the background plasma is modeled by a linear term which, absent $\phi_0$, causes relaxation, at rate $\nu$, to the background distribution function[12], $f_0$. $\nu$ may be interpreted as the rate of escape of trapped electrons. For convenience, $\varepsilon$ is re-expressed in terms of the nonlinear susceptibility, $\Xi$,

$$\varepsilon = 1 - \Xi\big/(k\lambda_D)^2. \qquad (2)$$

As $\phi$ and $\nu \to 0$, with $\nu/\omega_b << 1$, $\Xi \to \mathrm{Re}\,\Xi_0(\nu/v_e)$,

$$\Xi_0(x) = Z'\big(x/\sqrt{2}\big)\big/2, \qquad (3)$$

for Gaussian $f_0$. $Z$ is the plasma dispersion function[13] and v the wave's phase speed, $\omega/k$. In this limit, $\varepsilon \to \varepsilon_0$[14], $\varepsilon_0(k,\omega) = 1 - \mathrm{Re}\,\Xi_0(\nu/v_e)\big/(k\lambda_D)^2$. To lowest order[15] in $\phi$, the correction to $\mathrm{Re}\,\Xi$ is given by Eq. (48) of R&R (assuming that $\nu/\omega_b << 1$),

$$\mathrm{Re}\,\Xi = \mathrm{Re}\,\Xi_0(\nu/v_e) - 1.76\,f_0''(\nu/v_e)\sqrt{e\phi/T_e}\,. \qquad (4)$$

$f_0$ is normalized so that it integrates to unity, and it is evaluated in the plasma rest frame, *e.g.*, in thermal equilibrium, $f_0(x)=\exp\big(-x^2/2\big)\big/\sqrt{2\pi}$. The $\sqrt{\phi}$ term in Eq. (4) has been shown[10] to be a quantitatively accurate approximation to the exact $\mathrm{Re}\,\Xi$ for $e\phi/T_e$ as large as 0.5, for v in the range $2.0 < \mathrm{v}/\mathrm{v}_e < 3.5$.

The imaginary part of $\Xi$ has[16] a trapping-reduced Landau damping-like part, the term with the $f_0''(\mathrm{v})$ factor in Eq. (5), and a residual part, the second term on the RHS of Eq. (5), which is independent of $\phi$ (again in the limit $\mathrm{v}/\omega_b \ll 1$),

$$\mathrm{Im}\,\Xi = 6.17(\mathrm{v}/\omega_b)f_0'(\mathrm{v}/\mathrm{v}_e) + (\mathrm{v}/k\mathrm{v}_e)\Delta(\mathrm{v}/\mathrm{v}_e). \tag{5}$$

$\Delta$ is a functional of $f_0$ given by Eq. (70) in R&R

### 1. Relevance of residual damping

In Fig. 1, $|\Delta|/k\lambda_D$ is shown for a $k\lambda_D$ range relevant to TPMI. In this range, $\Delta < 0$. While $k$ and v are independent variables in Eq. (5), when $\mathrm{Im}\,\Xi$ is used to evaluate a small amplitude LW's damping, $k$ and v are approximately related by $\varepsilon_0(k,\omega)=0$. The relative importance of the Landau damping-like and residual parts is determined by $\omega_b$. They are of equal magnitude on the curve shown in Fig. 2.

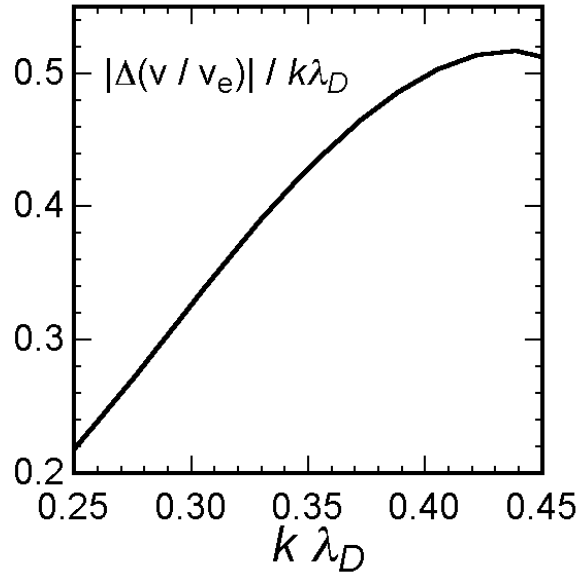

**FIG. 1. Normalized residual of** $\mathrm{Im}\,\Xi$. $k$ **is related to v through the Langmuir wave branch of Vlasov's dispersion relation,** $\varepsilon_0(k,\omega)=0$**: for the range of $k$ shown,** $\omega/\omega_p \approx 1+(3/2)(k\lambda_D)^2$**, the Bohm-Gross dispersion relation.**

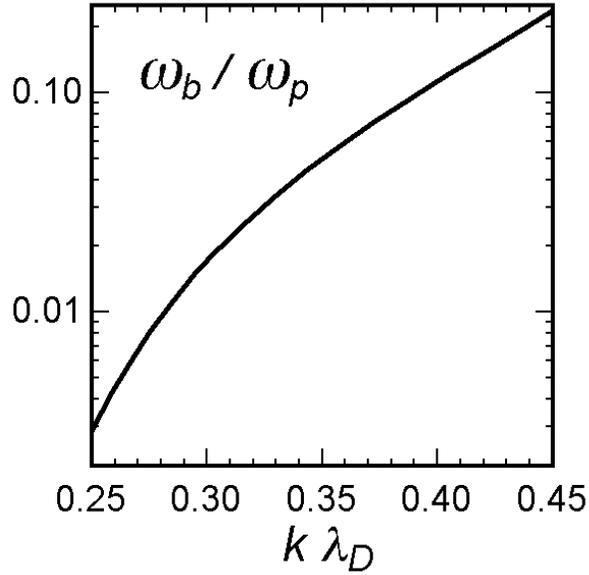

**FIG. 2. Bounce frequency at which the trapping-reduced Landau damping part of the total damping equals the residual damping part. Above this curve, residual damping dominates.**

Since in the above $k\lambda_D$ regime, $f_0'(v/v_e)$ and $\Delta$ are negative, $\left|\mathrm{Im}\,\Xi\right| > \left|(v/kv_e)\Delta(v/v_e)\right|$, and trapping may not lead to reduced damping. Once $\omega_b$ is above the value implied by

the curve in Fig. 2, the condition that trapping leads to reduced damping, $|\text{Im}\,\Xi| < |\text{Im}\,\Xi_0|$, may be approximated by

$$\nu/\omega_p < k\lambda_D \,\text{Im}\,\Xi_0(\nu/\nu_e)/\Delta(\nu/\nu_e). \tag{6}$$

The graph of the RHS of Eq. (6) is shown in Fig. 3. For given $k\lambda_D$, the ordinate of this curve gives the value of $\nu/\omega_p$ at which the trapped particle residual damping equals standard Landau damping. This residual part of the LW damping rate[17], to excellent approximation, is given by $1.5\nu$ in this $k\lambda_D$ regime.

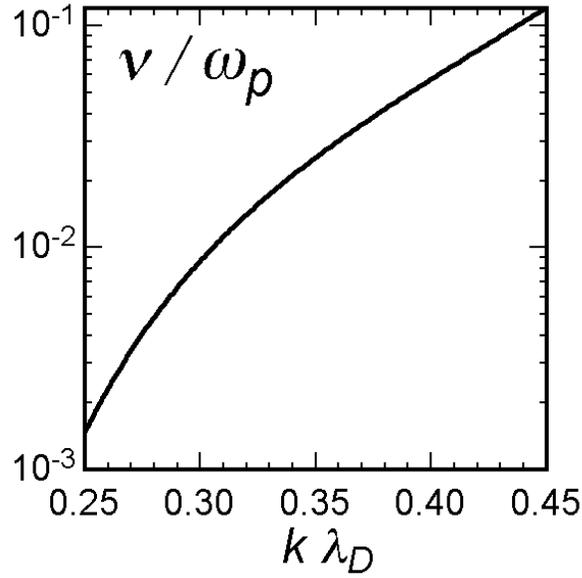

**FIG. 3. For values of trapped electron escape rate, $\nu$, below this curve, reduction of damping due to trapping is possible.**

### 2. Non-perturbative frequency shift evaluation

If, for given k and $\phi$, a resonance is possible, i.e., $\text{Re}(\varepsilon) = 0$ for some value of v, and if $v/\omega_b \ll 1$, conditions will be determined (see section III.C.1) such that there are solutions of (1) with $\text{Re}\,\varepsilon \approx 0$. Eqs. (2) and (4) then imply, to lowest order in $v/\omega_b$,

$$(k\lambda_D)^2 = \text{Re}\,\Xi_0(v/v_e) - 1.76 f_0''(v/v_e)\sqrt{e\phi/T_e}\,. \tag{7}$$

One may use Eq. (7) to simply graph $k$ as a function of v for given $\phi$, and invert this graph to determine $v(k, \phi)$ and hence an approximation to the nonlinear plasma mode dispersion relation[18]. Since (4) may be valid for physically large values of $\phi$, one might shortchange its content by solving Eq. (7) for v perturbatively in $\sqrt{\phi}$. For example, since $\text{Re}\,\Xi_0$ has a maximum value of $(0.534...)^2$ at $v/v_e \approx 2.12 \equiv v_*/v_e$, where the linear electron-acoustic[19] and Langmuir wave branches of the dispersion relation meet, Eq. (7) has no $\phi = 0$ solutions if $k > 0.534$. If $\phi \neq 0$, the maximum value of $k$ for which there is a resonance must decrease since[20] $f_0'' > 0$. Yet the expansion of v to lowest order in $\sqrt{\phi}$ gives a finite result for all $k < 0.534$, as shown in the next section. This loss of resonance phenomenon is well known, though it is often conflated with wave breaking[21][22].

## B. Perturbative evaluation of trapped particle frequency shift

The perturbative solution of (7) is reviewed[23] and compared with the results of Morales and O'Neil (M&O), Ref. 7. Denote an exact solution of (7) by $v(k, \phi)$. Let $v_0 = v(k, 0)$,

$$(k/k_D)^2 = \text{Re}\,\Xi_0(v_0/v_e), \tag{8}$$

and $v(k, \phi) = v_0 + \delta v(k, \phi)$. Substitution into Eq. (7) implies, to lowest order in $\omega_b$,

$$\Delta\omega_{\mathrm{TP}} \equiv k\delta v = 1.76\,\omega_b f_0''(v_0/v_e)\big/\mathrm{Re}\,\Xi_0'(v_0/v_e). \tag{9}$$

As $v_0 \to v_*$, $\Xi_0' \to 0$, and this result breaks down. Another way to say this is that at the loss of linear resonance, the perturbative expansion fails. Similarly for given $k$, there is a maximum value of $\phi$ beyond which nonlinear resonance is not possible, and as this value is approached, Eq. (9) must again fail. On the LW branch, Fig. 4 shows contours of non-perturbative $\Delta\omega_{\mathrm{TP}}$, as determined by Eq. (7), normalized to that given by Eq. (9). Note the progressive failure of $\Delta\omega_{\mathrm{TP}}$'s approximate bounce frequency scaling, as the loss of resonance boundary is approached (the last, unlabeled curve)

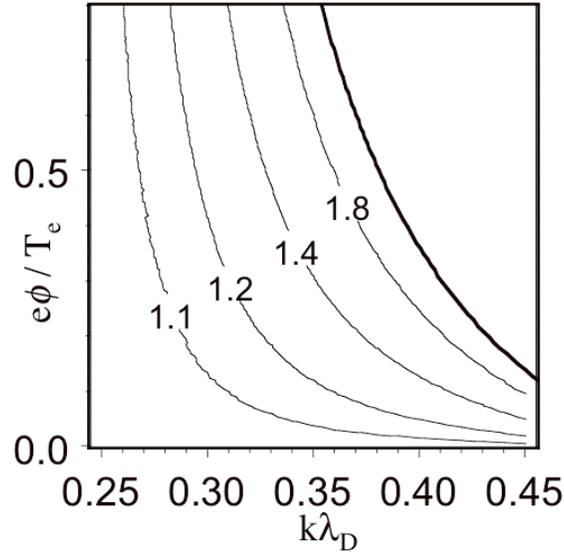

**FIG. 4 Contours of the non-perturbative frequency shift, as determined by Eq. (7), normalized to the first term in its bounce frequency expansion, given by Eq. (9). The $k$ axis coincides with the unit contour. The unlabeled upper curve is the loss of resonance boundary.**

In spite of its demonstrated limitation, the analytical bounce frequency approximation to $\Delta\omega_{TP}(k\lambda_D)$, Eq. (9), is useful, and its graph is shown in Fig. 5. It may be gleaned from Fig. 4 that the very rapid rise with increasing $k$, as shown in Fig. 5, is an underestimate.

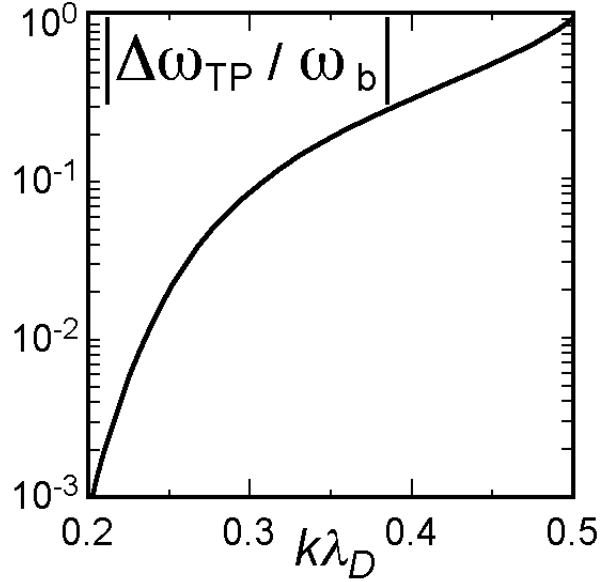

**FIG. 5. The magnitude of the (negative) trapped particle frequency shift, normalized to the electron bounce frequency, $\Delta\omega_{TP}/\omega_b$ as given by Eq. (9), with $k\lambda_D$, related to $v_0/v_e$ by the Langmuir wave branch of Eq. (8).**

The small $k\lambda_D$ limit of Eq. (9) is obtained using $\Xi_0'(x) \rightarrow -2/x^3$ for large $x$:

$$\Delta\omega_{TP} \xrightarrow[k \rightarrow 0]{} -(1.76/2)(v_0/v_e)^3 f_0''(v_0/v_e)\omega_b. \qquad (10)$$

The trapped particle frequency shift of M&O, $\Delta\omega_{MO}$, is almost the same: use the small-$k\lambda_D$-valid replacements, $\omega_p\partial\varepsilon/\partial\omega \rightarrow 2$ and $k \rightarrow \omega_p/v_0$, in Eq. (8) of M&O, to obtain

$$\Delta\omega_{MO} = -(1.63/2)(v_0/v_e)^3 f_0''(v_0/v_e)\omega_b. \qquad (11)$$

Eq. (9), for $v/v_e > 2.5$, yields a smaller frequency shift (by as much as a factor of 2 for $v/v_e \approx 3.2$ ($k\lambda_D \approx 0.37$)), than given by Eq. (10). This naïve extrapolation was not intended by M&O. Another extrapolation may be obtained by substitution of the exact relation, $(k\lambda_D)^3 \omega_p \partial \varepsilon_0 / \partial \omega = -\Xi'_0(v_0/v_e)$, into Eq. (8) of M&O. This recovers our Eq. (9), except for a factor of 1.63 compared to 1.76.

### C. Comparison with the ponderomotive frequency shift

The ponderomotive force induced Langmuir wave frequency shift[24], $\Delta\omega_{\text{pmf}}$,

$\Delta\omega_{\text{pmf}} = 0.5\omega_p \, \delta n/n = -\omega_p (k\lambda_D)^2 |e\phi/T_e|^2 / 8$, may also lead to modulational instability.

Equate $\Delta\omega_{\text{TP}}$, given by Eq. (9), to $\Delta\omega_{\text{pmf}}$ to obtain

$$(e\phi/T_e)^{3/2} = -14.1 f_0''(v/v_e) / \left[ k\lambda_D \Xi'_0(v/v_e) \right]. \tag{12}$$

For values of $\phi$ above the graph of this relation, shown as the solid curve in Fig. 6, the ponderomotive effect dominates[25]. However, $\Delta\omega_{\text{TP}}$ as given by Eq. (9) is not meaningful[26] if $\phi$ is close to or beyond, for given $k$, its loss of resonance value. This loss of resonance amplitude, as inferred from Eq. (7), is shown as the dashed curve in Fig. 6. Inside the region[27] labeled "$\Delta\omega_{\text{TP}} > \Delta\omega_{\text{pmf}}$", the trapped particle frequency shift is dominant. However, given the sharp rise of the solid curve, one expects that trapped particle effects will dominate at least for $\phi$ somewhat above the dashed curve, except that in this region, there are no traveling wave solutions to the Vlasov equation[28].

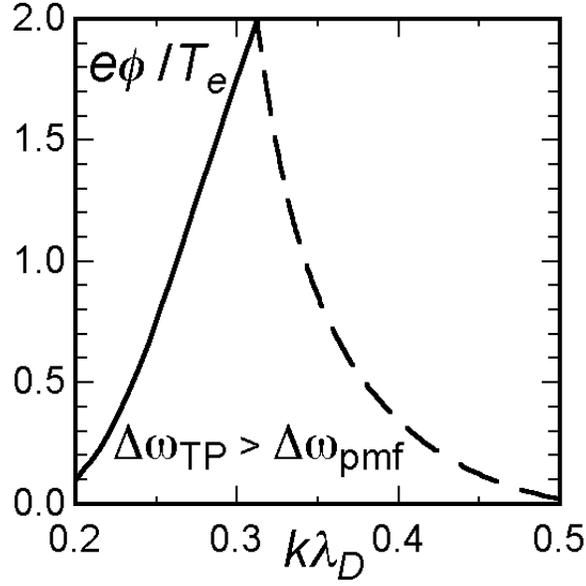

**FIG. 6. The trapped particle frequency shift has a larger magnitude than the ponderomotive force induced shift in the region bounded by the graph of Eq. (12) (solid curve), and the loss of resonance boundary (dashed curve) as determined by Eq. (7).**

## III. TRAPPED PARTICLE MODULATIONAL INSTABILITY

In this section, the trapped particle modulational instability (TPMI) growth rate, $\gamma_{\text{TPMI}}$, is determined and compared with the LDI growth rate, $\gamma_{\text{LDI}}$. Although damping rates are a critical part of a threshold analysis, it is simpler to first compare the growth rates *absent damping*. Electron thermal units are employed in the following: $k \to k/k_D$, $v \to v/v_e$ $(\omega \to \omega / \omega_p)$ and $\phi \to e\phi / T_e$, although for clarity, sometimes manifestly dimensionless expressions are used.

## A. The modulation model

Let a finite amplitude Langmuir traveling wave, Re $\phi_{res} \exp[i(k_z z - \omega t)]$, be an equilibrium solution[29] of the Vlasov equation[30]. $k$, $\omega$ and $\phi_{res}$, are related by the resonance condition

$$\varepsilon(k, \omega, \phi_{res}) = 0 . \tag{13}$$

$\varepsilon$ is approximately given by Eqs. (2), and (4) so that Eq. (7) is satisfied. Assume that departures from this equilibrium vary slowly in time[31] and space so that the modulation model,

$$\left[ \varepsilon + i \left( \frac{\partial \varepsilon_0}{\partial \omega} \frac{\partial}{\partial t} - \frac{\partial \varepsilon_0}{\partial k_z} \frac{\partial}{\partial z} \right) - \frac{1}{2} \frac{\partial^2 \varepsilon_0}{\partial k_x^2} \left( \frac{\partial^2}{\partial x^2} + \frac{\partial^2}{\partial y^2} \right) - \frac{1}{2} \frac{\partial^2 \varepsilon_0}{\partial k_z^2} \frac{\partial^2}{\partial z^2} \right] \phi = 0 , \tag{14}$$

may be useful[32]. The appearance of $\varepsilon$ and $\varepsilon_0$, both real, in Eq. (14) is explained below. Estimates based on the modulational instability show that $\partial / \partial t \sim \sqrt{\phi}$, so that the omitted $\partial^2 / \partial t^2$ term is assumed to be a correction for small $\phi$. Similarly, to lowest order in $\phi$, *the coefficients of all the derivative terms* in Eq. (14) have been evaluated using $\varepsilon(k, \omega) = \varepsilon_0(k, \omega)$. Eq. (14) models slowly varying departures from finite amplitude traveling wave solutions to the Vlasov equation, *i.e.*, BGK modes[33], not departures from (linear) traveling waves satisfying Landau's dispersion relation[34].

After derivatives of $\varepsilon_0$ are taken, $k$ and $\omega$ are related by $\varepsilon_0 = 0$. When evaluated at

$\mathbf{k}_0 = (0,0,k_z)$, $(\partial \varepsilon_0 / \partial k_x)_\omega = (\partial \varepsilon_0 / \partial k_y)_\omega = 0$, so that there are no $\partial / \partial x$ or $\partial / \partial y$ terms in Eq.

(14), and $\partial^2 / \partial x^2$ and $\partial^2 / \partial y^2$ terms simply combine since $(\partial^2 \varepsilon_0 / \partial k_x^2)_\omega = (\partial^2 \varepsilon_0 / \partial k_y^2)_\omega$. Let a fluctuation, $\delta \phi$, $\phi = \phi_{res} + \delta \phi$, vary as $\delta \phi \sim \exp(i \mathbf{p} \cdot \mathbf{x})$. In addition to slow time variation, it is required that the total wavevector, $\mathbf{k}$, $\mathbf{k} = \mathbf{k}_0 + \mathbf{p}$, be confined to a small neighborhood of $\mathbf{k}_0$, to conform to the notion of a wavepacket. One sense of neighborhood is that $\varepsilon_0$ is well represented by the first few terms of its Taylor series. A generalized modulation model that removes this limitation is discussed in section III.A.4.

If $\delta \phi$ is unstable, the maximum TPMI growth rate only depends[8] on $\Delta \omega_{TP}$. However,

the locus of $\mathbf{p}$ at which the maximum is attained depends on the diffraction coefficients,

and the frequency of the unstable mode depends on the group velocity. In order to understand these aspects of the instability, the group velocity and diffraction coefficients are calculated in the kinetic (*i.e.*, not necessarily small $k \lambda_D$) regime.

### 1. Group velocity

Since $\varepsilon_0$, is naturally expressed as a function of the independent variables $k$ and v,

evaluation of the derivative coefficients in Eq. (14), which require that $k$ and $\omega$ are independent, is facilitated by use of the identities $(\partial / \partial k)_\omega = (\partial / \partial k)_v - (v/k)(\partial / \partial v)_k$ and

$(\partial / \partial \omega)_k = (1/k)(\partial / \partial v)_k$. They imply, $(\partial \varepsilon_0 / \partial \omega)_k = -\text{Re} \, \Xi_0'(v)/k^3$ and

$(\partial\varepsilon_0/\partial k)_\omega = \mathrm{Re}\big[2\Xi_0(v) + v\Xi_0'(v)\big]\big/k^3$, from which the group velocity,

$v_G = -(\partial\varepsilon_0/\partial k_z)\big/(\partial\varepsilon_0/\partial\omega)$ may be evaluated[35]. It is graphed in Fig. 7.

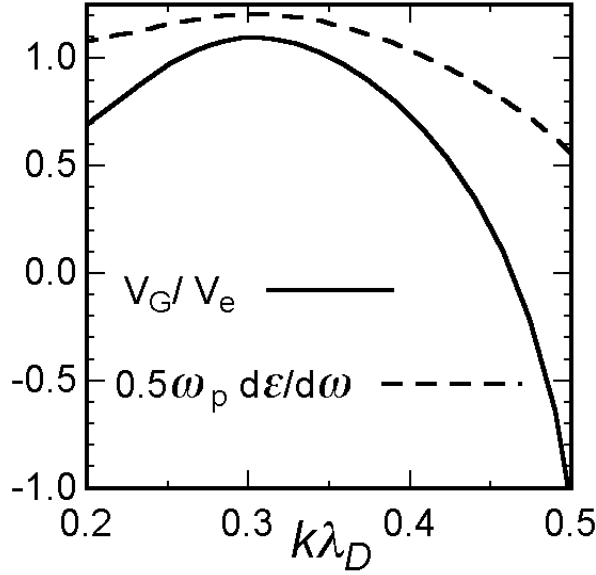

FIG. 7. The Langmuir wave group velocity (solid curve) begins to depart significantly from its small $k\lambda_D$ estimate, $3v_e k\lambda_D$, for $k\lambda_D > 0.3$. For reference, normalized $\partial\varepsilon_0/\partial\omega$ is also shown (dashed curve).

*2. Dispersion*

Eq. (14) may be re-written as:

$$i(\partial/\partial t + v_G\,\partial/\partial z)\phi = \big[\Delta\omega_{TP}(|\phi|) - \Delta\omega_{TP}(|\phi_{res}|)\big]\phi - D_\perp\nabla_\perp^2\phi - D_z\,\partial^2\phi\big/\partial z^2 \,. \qquad (15)$$

$\Delta\omega_{TP}$ is given by Eq. (9) and

$$\nabla_\perp^2 = \frac{\partial^2}{\partial x^2} + \frac{\partial^2}{\partial y^2}, \ D_\perp = -\frac{1}{2}\frac{\partial^2 \varepsilon_0 / \partial k_x^2}{\partial \varepsilon_0 / \partial \omega}, \ D_z = -\frac{1}{2}\frac{\partial^2 \varepsilon_0 / \partial k_z^2}{\partial \varepsilon_0 / \partial \omega}. \tag{16}$$

It follows from $(\partial \varepsilon_0 / \partial k_x)_\omega = (k_x / k)(\partial \varepsilon_0 / \partial k)_\omega$ that

$$D_\perp = v_G / 2k, \tag{17}$$

at $\mathbf{k} = \mathbf{k}_0$, while

$$\left( \frac{\partial^2 \varepsilon_0}{\partial k_z^2} \right)_\omega = -\frac{1}{k^4}\left( 6 + 6v\frac{d}{dv} + v^2\frac{d^2}{dv^2} \right) \mathrm{Re}\,\Xi_0. \tag{18}$$

$D_z$ goes through zero at $k\lambda_D \approx 0.33$, as seen in Fig. 8. $D_\perp$ has a qualitatively different appearance, going through zero at $k\lambda_D \approx 0.46$. Both approach the classical thermal equilibrium value of 3/2 as $k\lambda_D \to 0$.

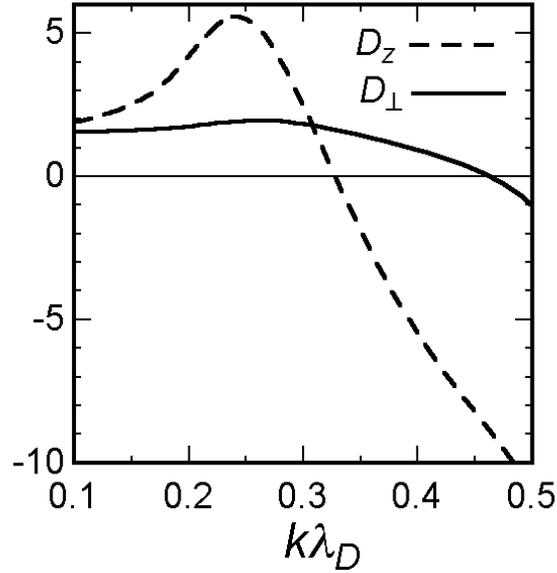

**FIG. 8. Diffraction coefficients go negative as $k\lambda_D$ increases: first $D_z$ then $D_\perp$.**

### 3. Modulational instability regimes

If the equilibrium solution of Eq. (15), $\phi = \phi_{res}$, is perturbed,

$$\phi = \phi_{res} + \delta\phi(t)\exp\left[i(\mathbf{p}_\perp \cdot \mathbf{x}_\perp + p_z z)\right],$$  (19)

with $\mathbf{p}_\perp = (p_x, p_y)$, $p_z$ real, and $\mathbf{x}_\perp = (x, y)$, then instability requires[8] positive frequency dispersion, $D$,

$$D = D_\perp p_\perp^2 + D_z p_z^2 > 0,$$  (20)

since $\partial(\Delta\omega_{TP})/\partial\phi < 0$. In the positive dispersion regime, $k_z\lambda_D < 0.33$, $D_\perp$ and $D_z$ are positive, instability may occur for any direction of $\mathbf{p} = (\mathbf{p}_\perp, p_z)$. In the mixed dispersion regime, $0.33 < k_z\lambda_D < 0.46$, $D_z$ is negative and quickly becomes larger in magnitude than $D_\perp$, so that unstable modes are elongated about the $z$ axis. In the negative dispersion regime, $k_z\lambda_D > 0.46$, instability is not possible. Let $\delta\phi(t) \sim \exp(\lambda t)$. In the frame of reference moving along the $z$ axis with velocity $v_G$, $\lambda$ satisfies the standard modulational dispersion relation, that follows from Eqs. (15) and (19):

$$\lambda^2 = D[-\Delta\omega_{TP}/2 - D].$$  (21)

If $0 < D < -\Delta\omega_{TP}/2$, then $\lambda$ is real, and it attains its maximum value of $|\Delta\omega_{TP}/4|$ at $D = |\Delta\omega_{TP}/4|$. The different dispersion regimes are illustrated in Figs. 9, 10 and 11, with the total wavevector, $\mathbf{k} = \mathbf{k}_0 + \mathbf{p}$, as independent variable.

Fig. 9 is the case $k_0\lambda_D = 0.30$. Because $1.8 \approx D_\perp < D_z \approx 2.5$, contours of $D$ (dashed curves) are elliptical, slightly elongated in the $k_x$ direction (note factor of 2 difference in $k_x$ and $k_z$ scales). The solid curves in Fig. 9, Fig.10 and Fig. 11 are discussed later.

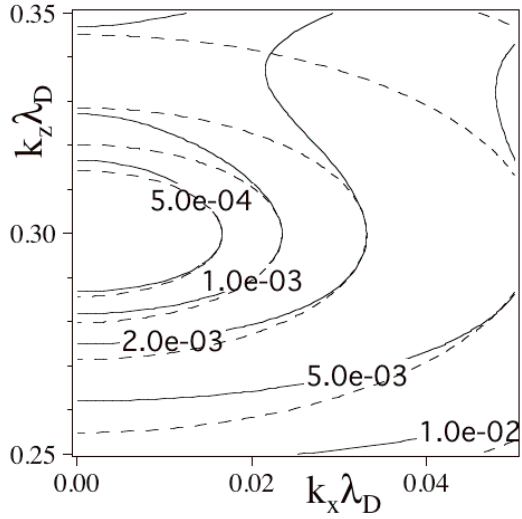

**FIG. 9. Contours (dashed curves) of frequency dispersion, *D*, for $k_0\lambda_D = 0.30$, as given by Eq. (20). Solid curves are contours of $\hat{D}_0$, Eq. (25), which is an exact evaluation of dispersion.**

This is the standard geometry for modulational instability, with, *e.g.*, stability for

$$D_\perp k_x^2 > |\Delta\omega_{TP}/2|. \qquad (22)$$

Fig.10 is the case $k_0\lambda_D = 0.345$, with $D_\perp \approx 1.5$ and $D_z \approx -1.5$, so that contours of *D* are hyperbolic. Only positive valued contours are shown as they determine TPMI.

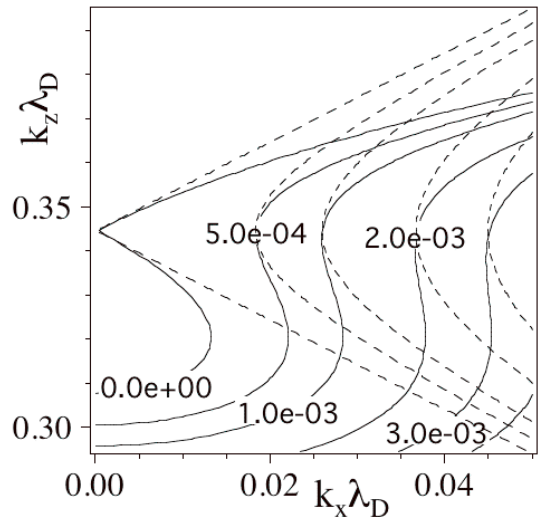

**FIG. 10. Same as Fig. 9, but** $k_0\lambda_D = 0.345$.

Fig. 11 illustrates the case $k_0\lambda_D = 0.398$, with $D_\perp \approx 1.0$ and $D_z \approx -5.0$, so that the contours of $D$ are stretched out more in the $k_x$ direction than in the case of Fig.10 (note the smaller range of $k_z$ in comparison with Fig.10).

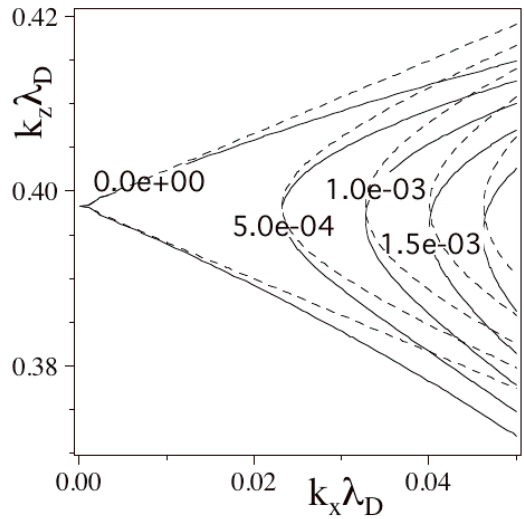

**FIG. 11.  Same as Fig. 9, but** $k_0 \lambda_D = 0.398$.

The dashed curves shown Fig.10 and Fig. 11 are superficially identical, except for a change of scale in the $k_z$ axis.  However, a scale change does not leave the physics invariant when a source with an independent length scale is added.  For example, if the source is due to backward stimulated scatter (BSRS) in a diffraction limited laser beam, the beam width is such a scale.  Also, solutions of (15) may sample Fourier modes far enough from $(0,0,k_z)$ such that the truncated Taylor series in Eq. (14) loses accuracy.

### 4. Generalized linear dispersive model

If $\varepsilon$ is expanded in deviations from $\omega$ but not $\mathbf{k}_0$, than instead of Eq. (14) one obtains

$$\left[ i \frac{\partial \varepsilon_0(\mathbf{k}_0, \omega)}{\partial \omega} \frac{\partial}{\partial t} + \varepsilon(\mathbf{k}_0 + \mathbf{p}, \omega) \right] \phi(\mathbf{p}, t) = 0 , \qquad (23)$$

where $\phi(\mathbf{p}, t)$ is the Fourier transform (FT) in space, but not time[36]. The dependence of $\varepsilon$ on $\phi$ is suppressed. At the TPMI threshold for a large damping rate, $|\Delta \omega_{\mathrm{TP}}|$ and hence $|\mathbf{p}|$ may be large enough to invalidate the expression for LW dispersion, Eq. (20), which follows from Eq. (14), while Eq. (23) may still be valid.  If the linear variation of $\varepsilon$ with $\mathbf{p}$ is extracted, so that the effect of finite group velocity is made explicit, then the remainder may be identified as generalized LW dispersion.  Instead of Eq. (15) one has:

$$i(\partial/\partial t + v_G \partial/\partial z)\phi = \left[\Delta\omega_{TP}(|\phi|) - \Delta\omega_{TP}(|\phi_{eq}|)\right]\phi + \hat{D}\phi .$$  (24)

$\hat{D}$ is a linear operator with the Fourier space representation

$$\hat{D}_0(\mathbf{p}) = -\left[\varepsilon_0(\mathbf{k}_0 + \mathbf{p}, \omega) - \varepsilon_0(\mathbf{k}_0, \omega) - p_z \cdot \partial\varepsilon_0(\mathbf{k}_0, \omega)/\partial k_z\right]/(\partial\varepsilon_0/\partial\omega) .$$  (25)

Note that $\varepsilon_0(\mathbf{k}_0, \omega) = 0$, but it is formally retained to emphasize the fact that were $\hat{D}_0$ expanded in $\mathbf{p}$, it would begin at second order and reduce to the expression given by Eq. (20). The solid curves in Fig. 9, Fig.10 and Fig. 11 are contours of $\hat{D}_0$. Departures from its second order expansion, the dashed curves, are apparent in Fig. 9 and Fig.10, and much less so in Fig. 11 because $k_z$ for the latter is well past the value at which $D_z$ changes sign. These two contour sets overlap for $\mathbf{p} \to \mathbf{0}$ ($\mathbf{k} \to \mathbf{k}_0$), as seen in Fig. 9, Fig.10 and Fig. 11.

## 5. Nonlinear dispersion

Values of $\phi$ of physical interest may be large enough that amplitude dependent diffraction coefficients must be considered. In the previous expressions for the linear diffraction coefficients, Eq. (16), replace $\varepsilon_0$ by $\varepsilon$, as determined by Eqs. (2) and (4), with $k$ and $v$ related by the nonlinear dispersion relation, Eq. (7). Note that Eq. (17) for $D_\perp$ still holds in the nonlinear regime. Fig. 12 and Fig. 13 show the resulting $D_z(k, \phi)$ and $D_\perp(k, \phi)$ respectively, with curves labeled by $e\phi/T_e$. The topmost in each of these figures has $\phi = 0$, reproducing the curves in Fig. 8.

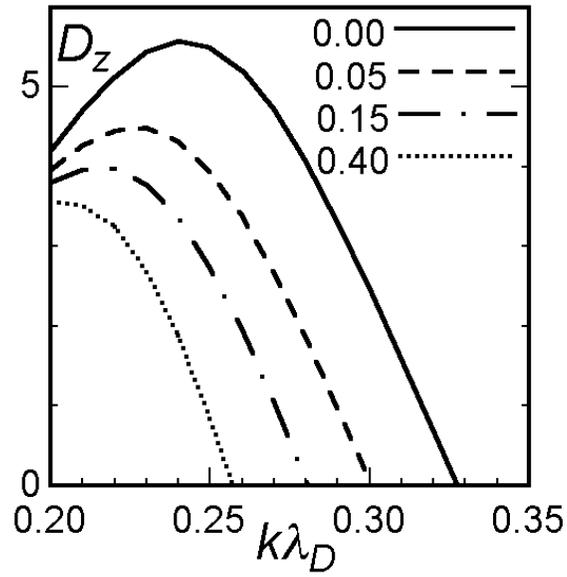

**FIG. 12.** Diffraction coefficient in wave propagation direction, $D_z$, parameterized by the dimensionless wave amplitude, $e\phi/T_e$. If $k\lambda_D$ is beyond the intercept with zero for given $\phi$, then $D_z < 0$, *e.g.*, for $e\phi/T_e = 0.05$, this is approximately the range $k\lambda_D > 0.3$.

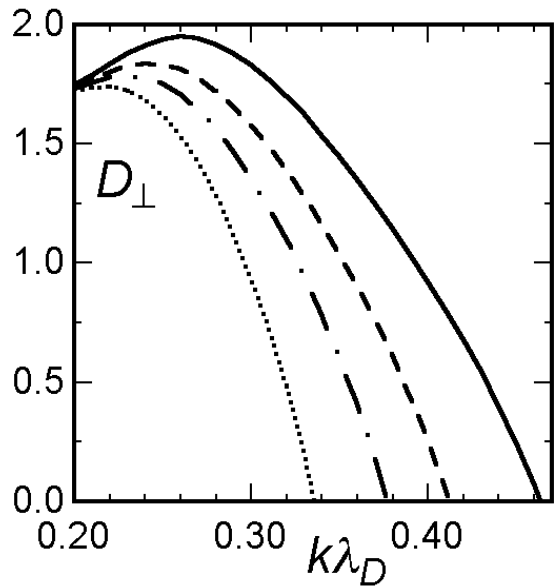

**FIG. 13. Diffraction coefficient, perpendicular to wave propagation direction, $D_\perp$, parameterized by the wave amplitude, $e\phi/T_e$. Curves are labeled by $e\phi/T_e$, as in Fig. 12.**

Note that for given $k\lambda_D$, there is a cutoff value of $e\phi/T_e$, above which the diffraction coefficient goes negative, *e.g.*, for $D_z$ and $k\lambda_D = 0.3$, this occurs at $e\phi/T_e \approx 0.05$, while for $D_\perp$ and $k\lambda_D = 0.38$, this occurs at $e\phi/T_e \approx 0.15$. Fig. 14 shows the cutoff wave amplitude for the two diffraction coefficients. Above the solid curve, both coefficients are negative, and TPMI is not possible. Note that this curve lies below the loss of resonance curve (see Fig. 15), and therefore is the stronger constraint on TPMI.

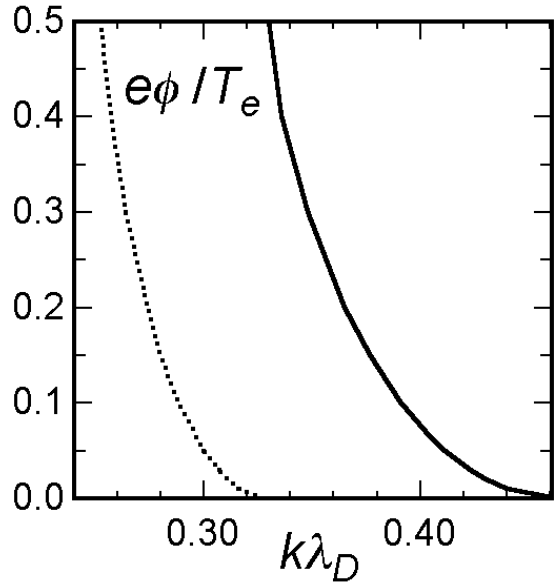

**FIG. 14. Amplitude cutoff for positive wave dispersion: above the dotted curve, $D_z < 0$. Above the solid curve, $D_\perp < 0$, positive dispersion is not possible, and neither is TPMI.**

Unlike the generalized linear dispersion model, Eq. (25), the nonlinear diffraction model has no such counterpart unless $\phi$'s amplitude varies slowly compared to its phase. Such a model is called for in a spatial region where one of the diffraction coefficients goes through zero, a not unlikely scenario when SRS convectively grows from thermal fluctuations.

## B. Comparison of TPMI and LDI growth rates absent dissipation

The TPMI growth rate, $\gamma_{TPMI} = |\Delta\omega_{TP}/4|$,[37] while the LDI growth rate, $\gamma_{LDI}$, for a LW with wavenumber $k >> k_*$, is well known and given (*in electron thermal units*) by[38]

$$\gamma_{LDI} = k\phi\sqrt{c_s k/8}, \qquad (26)$$

if $\gamma_{LDI}$ is small compared to the ion acoustic frequency, $kc_s$. Solution of the full dispersion relation[39] shows that ion inertia reduces the LDI growth rate by no more than 14%, compared to Eq. (26), for $k < 0.5$ and $\phi < 3.0$. This difference will be ignored here, allowing for simple analytic comparison with $\gamma_{TPMI}$. Equate $\gamma_{TPMI}$, with $\Delta\omega_{TP}$ given by Eq. (9), to $\gamma_{LDI}$, to obtain the value of $\phi$ below which TPMI has the larger growth rate:

$$\sqrt{c_s k\phi/8} = 0.44 f_0''(v)\Big/\text{Re}\,\Xi_0'(v), \qquad (27)$$

shown as the solid curve in Fig. 15, with $k$ and $v$ related by the Langmuir branch of Eq. (8).

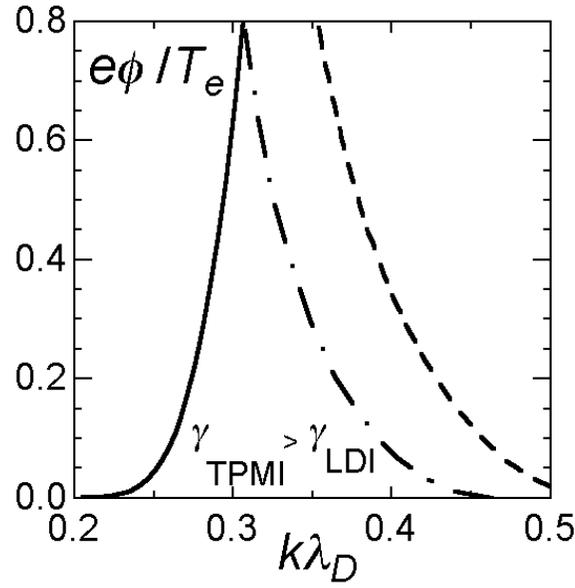

**FIG. 15. The trapped particle modulational instability has a larger growth rate than the Langmuir ion-acoustic decay instability for values of $\phi$ below the solid curve, the graph of Eq. (27), and below the dash-dot curve, the positive dispersion boundary (also shown as the solid curve in Fig. 14). Dashed curve is the loss of resonance boundary. The speed of sound has been assigned the value $v_e/50$.**

However, if $\phi$ is above the dash-dot curve, then TPMI cannot occur. The analysis of TPMI breaks down near this curve: as it is approached from below, $D_\perp \downarrow 0$, and maximum growth rate occurs at larger and larger values of the perturbation wavenumber, $p$, violating the assumed small $p$ limit.

### C. TPMI versus LDI damping thresholds

TPMI induced fluctuations have wavevectors close to $\mathbf{k}_0$, while, in contrast, the LDI daughter Langmuir wave tends to point[40] in the direction opposite to $\mathbf{k}_0$: the former may have its damping rate directly reduced by electron trapping, while the latter does not[41]. The combination of reduced damping and greater TPMI growth rate (see Fig.15), and the modification of the LDI frequency/wavenumber matching conditions due to the trapped particle frequency shift of the primary LW, is shown to lead to a rich structure of competition between these instabilities, as $k\lambda_D$ is varied. As a prelude to the TPMI threshold calculation, spatially uniform perturbations are analyzed.

### 1. Stability of damped/driven equilibria to spatially uniform perturbations

When dissipation and external potential, $\phi_0$, are retained, and only spatially uniform fluctuations are allowed, one obtains in place of Eq. (15),

$$i(d/dt + \gamma_{TP})\phi = \left[\Delta\omega_{TP}(|\phi|) - \Delta\omega_{TP}(|\phi_{res}|)\right]\phi + \phi_0/(\partial\varepsilon_0/\partial\omega). \qquad (28)$$

The trapped particle damping rate,

$$\gamma_{TP} = \mathrm{Im}\,\varepsilon/(\partial\varepsilon_0/\partial\omega), \qquad (29)$$

is determined by Eqs. (2) and (5). There may be as many as three[42] equilibrium solutions, $\phi_{eq}$, to Eqs. (1) and (28),

$$(\Delta\omega - i\gamma_{TP})\,\phi_{eq} + \phi_0/(\partial\varepsilon_0/\partial\omega) = 0. \qquad (30)$$

$\Delta\omega = \Delta\omega_{\text{TP}}(|\phi_{eq}|) - \Delta\omega_{\text{TP}}(|\phi_{res}|)$, the frequency departure from resonance. A condition is

now obtained such that two of these are close to resonance, $|\phi_{eq}| = |\phi_{res}| + \delta$, $|\delta|/|\phi_{res}| \ll 1$.

Since $\Delta\omega_{\text{TP}} \sim \sqrt{|\phi|}$, $\Delta\omega \approx 0.5\Delta\omega_{\text{TP}}\delta/|\phi_{eq}|$. Since $\delta$ is small, $\gamma_{\text{TP}}$ may be evaluated at

$|\phi| = |\phi_{res}|$,

$$\left(0.5\Delta\omega_{\text{TP}}\,\delta/|\phi_{eq}| - i\gamma_{\text{TP}}\right)\phi_{eq} = -\phi_0/(\partial\varepsilon_0/\partial\omega).\tag{31}$$

The phase of $\phi_{eq}$ may be determined once $\delta$ is found from

$$\left(0.5\Delta\omega_{\text{TP}}\delta\right)^2 + \gamma_{\text{TP}}^2|\phi_{eq}|^2 = |\phi_0|^2\Big/(\partial\varepsilon_0/\partial\omega)^2.\tag{32}$$

If $\gamma_{\text{TP}}|\phi_{eq}| < |\phi_0|/(\partial\varepsilon_0/\partial\omega)$, which simplifies to

$$\gamma_{\text{TP}}/\omega_p < 0.5|\phi_0|/|\phi_{eq}|\tag{33}$$

when $\partial\varepsilon_0/\partial\omega \to 2/\omega_p$, then Eq. (32) has two real solutions for $\delta$. Since

$0.5\Delta\omega_{\text{TP}}|\delta| < |\phi_0|/(\partial\varepsilon_0/\partial\omega)$, the condition $|\delta|/|\phi_{res}| \ll 1$ and Eq. (33) imply

$$2\gamma_{\text{TP}}/\omega_p < |\phi_0|/|\phi_{res}| \ll |\Delta\omega_{\text{TP}}|/\omega_p.\tag{34}$$

In this regime, the external potential does not much affect equilibrium amplitudes, but merely their phases, via Eq. (31).

Eq. (33) identifies a threshold value of $\phi_0$, below which there are no nearly resonant solutions to Eq. (28). This is illustrated in Fig. 16. The non-resonant solution (not shown) is typically much smaller, *e.g.*, for the $v/\omega_p = 0.001$ cases, by more than an order of magnitude near threshold.

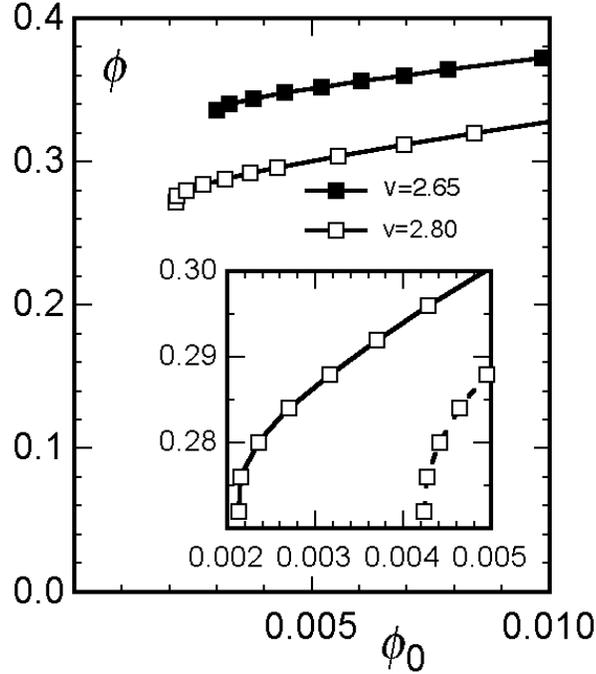

**FIG. 16[a15]. Larger of the two nearly resonant solutions to Eq. (30), for** $k\lambda_D = 0.4$ **and** $v/\omega_p = 0.001$**, labeled by their phase velocity,** $v/v_e$**. Inset shows threshold regime for two** $v/v_e = 2.8$ **cases: solid curve** $v/\omega_p = 0.001$**, and lower dashed curve** $v/\omega_p = 0.002$**.**

Linearize Eq. (28) about an equilibrium solution (not yet assuming any ordering of parameters), $\phi = \phi_{eq} + \theta$, to obtain

$$\left\{ \partial/\partial t + \frac{i}{4}\Delta\omega_{\text{TP}}\left(\left|\phi_{eq}\right|\right) + i\Delta\omega + \frac{3}{4}\gamma_{\text{TP}}\left(\left|\phi_{eq}\right|\right) \right\}\theta = \left(\frac{\gamma_{\text{TP}}}{4} - i\frac{\Delta\omega_{\text{TP}}}{4}\right)\theta^{*}. \qquad (35)$$

The constant part of $\gamma_{\text{TP}}$ is momentarily suppressed so that $\gamma_{\text{TP}} \propto 1/\sqrt{\phi}$. Let $\theta \sim \exp(\lambda t)$. Then,

$$[\lambda + 3\gamma_{TP}/4]^2 = \Delta\omega(|\Delta\omega_{TP}|/2 - \Delta\omega) + (\gamma_{TP}/4)^2. \tag{36}$$

If $|\delta|/|\phi_{res}| \ll 1$, then $|\Delta\omega| \ll |\Delta\omega_{TP}|$, and

$$[\lambda + 3\gamma_{TP}/4]^2 = \Delta\omega|\Delta\omega_{TP}|/2 + (\gamma_{TP}/4)^2. \tag{37}$$

Since $\Delta\omega < 0$ if $|\phi_{eq}| > |\phi_{res}|$, the larger of the two nearly resonant solutions is always stable.

### 2. Stability against spatially varying fluctuations

If a fluctuation with wavevector $\mathbf{p}$ is allowed, it can be shown that the dispersion relation is obtained from Eq. (36) by the substitution $\Delta\omega \to \Delta\omega + \hat{D}(\mathbf{p})$. Recall that $\hat{D}(\mathbf{p})$ is the generalized dispersion given by Eq. (25). As in the analysis of Eq. (21), the most unstable fluctuation is attained when $\Delta\omega + \hat{D}(\mathbf{p}) = |\Delta\omega_{TP}|/4$, and then $[\lambda + 3\gamma_{TP}/4]^2 = (|\Delta\omega_{TP}|/4)^2 + (\gamma_{TP}/4)^2$. At threshold ($\lambda = 0$),

$$|\Delta\omega_{TP}|/4 = \gamma_{TP}/\sqrt{2}. \tag{38}$$

The $1/\sqrt{2}$ factor arises from the inverse dependence of $\gamma_{TP}$ on $\sqrt{\phi}$. If $\gamma_{TP}$ were $\phi$ independent, then the threshold condition would have, instead, a factor of unity. In general, the factor interpolates between these two limits, when the complete expression for $\mathrm{Im}\,\varepsilon$ is used. For simplicity this distinction will be ignored for the purpose of comparison with the LDI threshold, and the factor of $\sqrt{2}$ in Eq. (38) is ignored. That understood, Eqs. (5), (9), (29) and (38) imply that at threshold for TPMI (in thermal units),

$$1.76\omega_b^2 f_0''(v) + 4v[\omega_b \Delta(v) + 6.17 k f_0'(v)] = 0 \ . \tag{39}$$

For the case $v/\omega_p = 0.003$, the solution is shown as the solid curve in Fig. 17. TPMI is possible above the solid curve and, when nonlinear dispersion is used, below the positive dispersion boundary, the dash-dot curve.

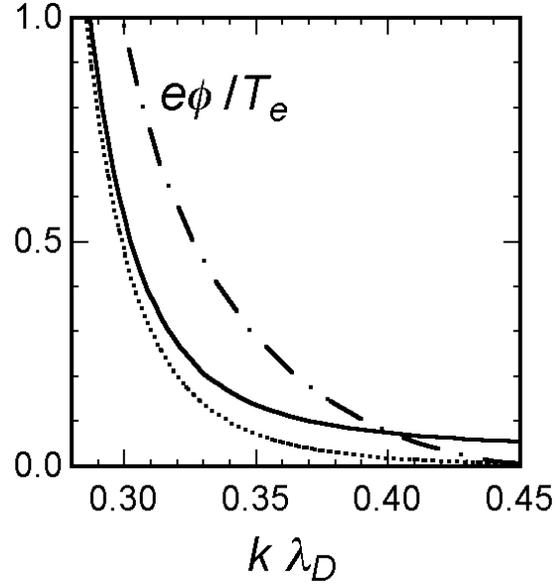

FIG. 17 TPMI  electrostatic potential amplitude threshold for trapped electron escape rate $v/\omega_p = 0.003$, solid curve.  Approximate threshold (dotted curve) obtained by omitting trapping reduced Landau damping (the $f_0'(v)$ term in Eq. (39)).  Upper, dash-dot curve, is again the Langmuir wave positive dispersion boundary.

If the $f_0'(v)$ term is ignored in Eq. (39), then

$$\sqrt{\phi} \approx -2.3 v (\Delta(v)/k)/f_0''(v), \tag{40}$$

whose graph is the dotted curve in Fig. 17.  If, in contrast, $\Delta$ is ignored, and only the Landau-damping-like -contribution to $\gamma_{TP}$ is retained, then

$$\phi \approx 14\nu, \tag{41}$$

is obtained for the TPMI threshold when $\nu^2 >> 1$.  This becomes accurate for larger $k\lambda_D$ as the threshold amplitude decreases.  The wavenumber independence of this estimate is due to the near balance between the increase of $\left|\Delta\omega_{TP}\right|$ with $k\lambda_D$ and the increase of Landau damping.

### 3. Reduction of LDI threshold by primary LW trapped particle frequency shift

First, the standard LDI threshold analysis is recalled.  Let the primary (daughter) LW have wavenumber $k$ ($p$).  At the threshold, based on the linear dispersion relation for the primary and daughter waves, $\gamma_{LDI}^2 = k^3 \phi^2 c_s / 8 = v_L(p) v_{ia}(k+p) c_s$, the product of the daughter wave damping rates, with $p = k - k^*$.  $v_{ia}$ is the ion acoustic damping rate coefficient.   In the $k >> k^*$ regime,

$$k^2 \phi^2 \approx 16 v_L(p) v_{ia}. \tag{42}$$

$v_L(k)$ is the standard LW Landau damping function, $v_L(k) = \sqrt{\pi/8}\left(1/k^3\right)\exp\left(-0.5\nu^2\right)$, with $k$ and $\nu$ again related by the linear dispersion relation. The magnitude of $k^*\lambda_D = 2c_s/3\nu_e$ is determined by the frequency of the daughter ion acoustic frequency.  The occurrence of the $k - k_*$ argument of Landau damping, in lieu of $k$, is significant only when Landau damping varies substantially over the range $k - k^*$ to $k$. For example, if $k^* \lambda_D = 0.01$ then $v_L(0.30)/v_L(0.29) \approx 1.3$, but $v_L(0.21)/v_L(0.20) \approx 2.7$.

Of possibly greater importance than the ion acoustic frequency induced wavenumber shift, is the change in the LDI wavenumber/frequency matching condition due to the primary LW trapped particle frequency shift. When this is taken into account, wavenumber and frequency matching imply

$$\omega(k, \phi) = \omega(k, 0) + \Delta\omega_{\mathrm{TP}}(k, \phi) = \omega(p, 0) + (k + p)c_s. \quad (43)$$

$\omega(k, \phi) = k\,v(k, \phi)$ is the LW solution of Eq. (7). For small $(k - p)$, $(k - p)\,d\omega/dk$ $\approx -\Delta\omega_{\mathrm{TP}}(k, \phi) + 2kc_s$, and the shifts are additive. As a practical matter note that the LW linear dispersion relation satisfies $d\omega/dk \approx 3k\lambda_D v_e$ to within 20% accuracy for $k\lambda_D < 0.37$, so that in this range,

$$(k - p)\lambda_D = k^*\lambda_D - \Delta\omega_{\mathrm{TP}}(k, \phi)/3k\,v_e. \quad (44)$$

Fig. 18 shows contours of $(k - p)\lambda_D$ as determined by Eqs. (7) and (43), for $k^*\lambda_D = 0.01$.

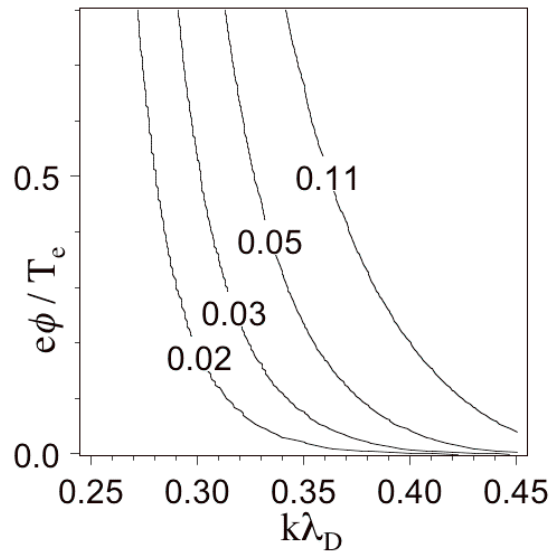

**FIG. 18 Contours of** $(k-p)\lambda_D$, **the LDI daughter LW wavenumber shift,**

**for** $k^*\lambda_D = 0.01$, **as a function of the primary LW's wavenumber,** $k\lambda_D$, **and potential,**

$e\phi/T_e$ **.**

Because of frequency matching, Eq. (43), the primary LW trapped particle frequency shift has, in effect, been transferred to the LDI daughter LW. The corresponding daughter LW Landau damping, $\nu_L(p)$, implied by Fig. 18, is shown in Fig. 19. For example, inspection of the $k$ axis in Fig. 19 shows that, for a primary LW with $k\lambda_D = 0.3$, the daughter LW Landau damping has the value $0.01\omega_p$. This is consistent with the fact that on this axis, $\phi = 0$, and so the daughter wave is at $p\lambda_D = 0.3 - k^*\lambda_D = 0.29$, and $\nu_L(0.29) \approx 0.01\omega_p$. Other points on this $0.01\omega_p$ contour in Fig. 19 may be determined from Fig. 18 as follows. Consider, $e.g.$, the $(k-p)\lambda_D = 0.05$ contour in Fig. 18. If $k\lambda_D = 0.34$, then $p\lambda_D \approx 0.29$, and $e\phi/T_e \approx 0.3$, which is consistent with the $0.01\omega_p$ contour of Fig. 19 going through the same point, $k\lambda_D = 0.34$ and $e\phi/T_e \approx 0.3$.

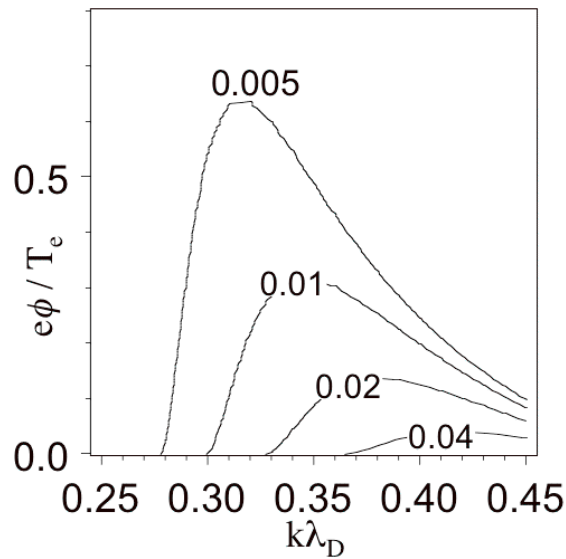

**FIG. 19 Contours of $\nu_L(p)/\omega_p$, LDI daughter LW's Landau damping,**

**for $k^*\lambda_D = 0.01$.**

Eq. (42) now gives the LDI threshold amplitude, whose contours are shown in Fig. 20

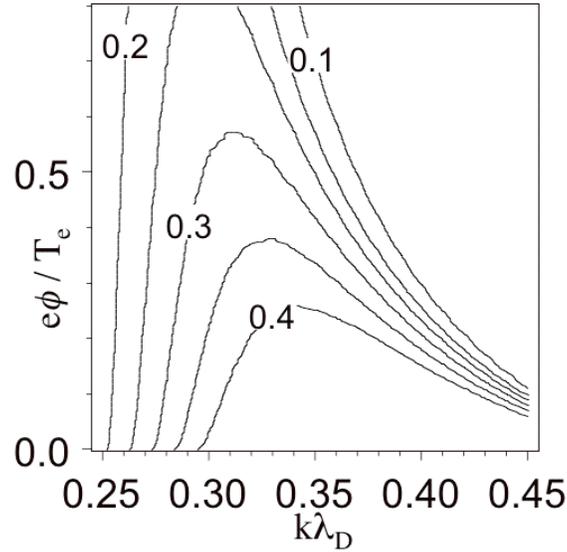

**FIG. 20 LDI potential amplitude threshold contours, normalized to $T_e/e$, for**

$\nu_{ia} = 0.1$ **and** $k^*\lambda_D = 0.01$.

for $\nu_{ia} = 0.1$. When this condition is made self-consistent,

$$\left(k\lambda_D\, e\phi/T_e\right)^2 = 16\nu_{ia}\nu_L\left[p(k\lambda_D, e\phi/T_e)\right].\tag{45}$$

It is illustrated by the following example: a horizontal line drawn across Fig. 20 at $\phi = 0.3$ intersects the 0.3 contour at $k \approx 0.29$ and $k \approx 0.37$. Repeating this exercise for all $\phi$ yields the linear LDI threshold, modified by the trapped particle frequency shift of the primary LW.  It is shown as the solid (dash-dot) curve, for $v_{ia} = 0.1$ ($v_{ia} = 0.025$), in Fig. 21.

If the trapped particle frequency shift of the primary LW is ignored, then the dotted curve in Fig. 20 is obtained for $v_{ia} = 0.1$.

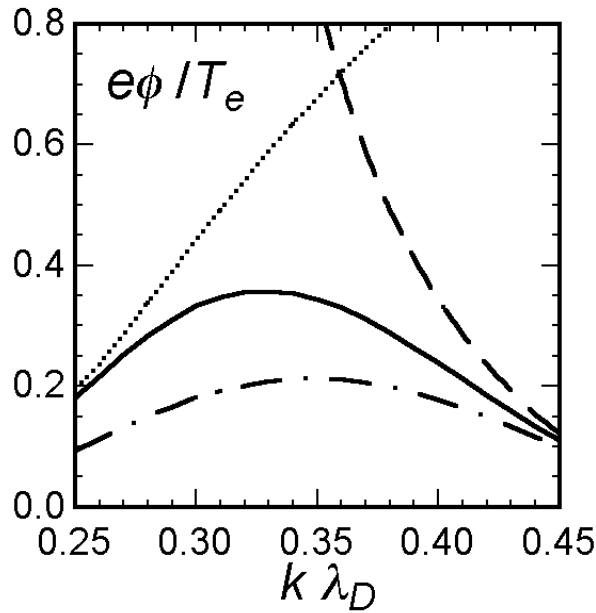

**FIG. 21 LDI threshold amplitude, for $v_{ia} = 0.025$, dash-dot, and $v_{ia} = 0.1$, solid, curve, self-consistently taking into account the trapped particle frequency shift of the primary LW.  If this shift is ignored, then the standard threshold is recovered, shown by the dotted curve.  For all cases, $k^* \lambda_D = 0.01$.  The dashed curve is LW loss of resonance.**

*4. Comparison of thresholds*

Comparison of Fig. 17 and Fig. 21 shows the TPMI and LDI threshold curves intersecting at roughly $k\lambda_D = 0.3$, where significant departures of the LDI threshold due to finite $\Delta\omega_{TP}(k,\phi)$ effects become significant. The standard LDI threshold scaling result, $\phi \propto \sqrt{v_{ia}}$, approximately valid for $k\lambda_D < 0.32$ (see Fig. 21), and the simplified TPMI threshold result, Eq. (40), imply that the LDI/TPMI threshold crossover depends on the fundamental damping coefficients only in the combination $(v/\omega_p)^4/v_{ia}$.

The thresholds are compared graphically in Fig. 22 for $v_{ia} = 0.1$, and various values of $v$.

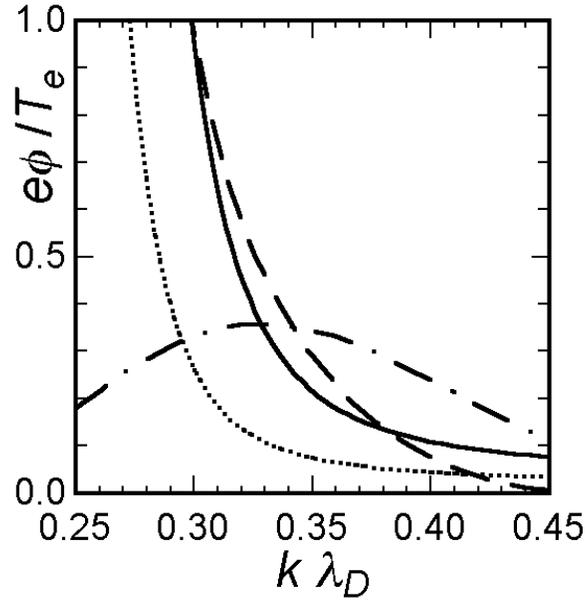

**FIG. 22 TPMI threshold for $v/\omega_p = 0.002$ (dotted curve), and $v/\omega_p = 0.004$ (solid curve). LDI threshold for $v_{ia} = 0.1$, and $k^*\lambda_D = 0.01$, dash-dot curve. LW positive dispersion boundary, dashed curve.**

Note that the TPMI threshold amplitude must lie below the positive dispersion boundary, the dashed curve. For $\nu/\omega_p \approx 0.005$, the threshold lies above this boundary except at $k\lambda_D \approx 0.34$, where the two curves meet. For greater values of $\nu$, TPMI is not possible because of the decrease of diffraction coefficients, eventually becoming negative, with increasing $\phi$.

## D. Langmuir wave self-focusing estimates and implications for an actual experimental beam

In Ref. [9], backscatter SRS and its daughter LW are observed in a large aspect ratio, near diffraction limited laser beam, propagating in the "$z$" direction, in a very homogeneous background plasma. The most basic estimate as to the relevance of possible electron trapping effects relates to the validity of the strongly trapped condition, $\nu/\omega_b \ll 1$, namely that an electron oscillates many times in the LW before escaping. Based on measured reflectivity values, and experimental parameters, it will be shown that this inequality is easily satisfied.

### 1. Strong trapping made plausible

While there are various processes which may contribute to the rate of escape of trapped electrons, $\nu$, the only one considered here is electron advection across the LW diameter, assumed comparable to the beam waist (beam diameter at best focus), $\approx F\lambda_0$, where $F$ is the optic f/#, and $\lambda_0$ the laser wavelength. Dimensional analysis then leads to $\nu = A \, \mathrm{v}_e / F\lambda_0$. If the coefficient, $A$, is set to unity, then

$$2\pi F \, \nu / \omega_p = (\mathrm{v}_e / c)\sqrt{n_c / n_e} \,. \tag{46}$$

Nominal experimental parameters ($T_e = 500\,\mathrm{eV}$, $F = 4.5$, $n_c / n_e = 25$) then imply $\nu / \omega_p \approx 0.005$.

One estimate of the electron bounce frequency is obtained by first omitting diffraction from SRS dynamics[43], to obtain, at low density,

$$dE_{SRS}/dt < k^2 \mathrm{v}_{osc}\, \phi / 4 \,. \tag{47}$$

This is in the frame of reference moving at the scattered light group velocity, $\approx c$, the speed of light. $E_{SRS}(E_0)$ is the scattered (laser) light temporal and spatial envelope. $\mathrm{v}_{osc}$ is the electron oscillating speed in $E_0$ so that $E_{SRS}/\mathrm{v}_{osc} \propto \sqrt{R}$, with $R$ the SRS reflectivity. The inequality sign is introduced because this result will now be used to estimate magnitudes and because omitted diffraction tends to weaken SRS spatial gain. Estimate $t = L/c$, where $L$ is the plasma length over which $E_{SRS}$ is near its peak value, assumed large compared to its value at the far end of the speckle[44]. It follows that

$$(\omega_b / \omega_p)^2 > (2/\pi)(\lambda_0 / L)\sqrt{R\, n_c / n_e} \,. \tag{48}$$

Since $L$ is at most a speckle length, $\approx 7F^2\lambda_0$, and[45] $R > 0.01$ in the LDI regime, it follows that $\omega_b / \omega_p > 0.05$, which is 10 time larger than the value of $\nu / \omega_p$ obtained from dimensional analysis, Eq (46). Given $\lambda_0 = 0.53\,\mu m$, it follows that $1/\omega_b < 0.03\,\mathrm{ps}$, and any other time scale, such as variability of $\phi_0$ (variability of $R$) must correspondingly be sub-sub picosecond to break the strong bounce frequency ansatz. If $k\lambda_D = 0.3$, then this bounce frequency estimate implies that $e\phi/T_e > 0.025$, the experimentally estimated lower bound[9].

## *2. Instability transition and co-existence*

A qualitative estimate of the wavenumber interval in which the TPMI and LDI thresholds cross may be obtained, somewhat arbitrarily, by insisting that $\nu$ is no smaller than, *e.g.*, 1/5 of the nominal value given by Eq. (46), $0.001\omega_p$. For this value of $\nu$ it can be shown that the thresholds cross at $k\lambda_D \approx 0.275$, while at the largest value $\nu$ can attain and still allow TPMI, $0.005\omega_p$, the thresholds cross at $k\lambda_D \approx 0.34$ for $\nu_{ia} = 0.1$.

Even though thresholds may cross at a particular $k$, it does not follow that there is an associated gradual transition from LDI to TPMI activity in a particular experiment. There may be an instability gap. For example, if the largest value attained by $e\phi/T_e$ is 0.3, then in the range $0.29 < k\lambda_D < 0.32$, both LDI and TPMI are below threshold for $\nu/\omega_p = 0.003$ and $\nu_{ia} = 0.1$.

It is also possible that the two instabilities may co-exist in the same region of space if the primary LW amplitude is above the LDI threshold and below the TPMI threshold. Let LDI be strong enough so that a second cascade step occurs. The third daughter LW's (counting the primary as first) phase velocity is typically well within a trapping width of the primary since $\left|\Delta\omega_{\text{TP}}\right|$ is small compared to $\omega_b$ (*e.g.*, if $k\lambda_D < 0.35$ then

$\left|\Delta\omega_{\text{TP}}\right| < 0.2\omega_b$, see Fig. 5). Then the first and third LWs act together with regard to self-

focusing, possibly exceeding the TPMI threshold.

### 3. Finite beam constraint

The most unstable TPMI mode, with wavenumber $p_\perp$ in the *x-y* plane, satisfies

$D_\perp p_\perp^2 \lambda_D^2 = \left|\Delta\omega_{\text{TP}}/4\omega_p\right|$. Assume that the primary LW's $k\lambda_D \approx 0.35$, so that $D_\perp \approx 0.5$ for

wave amplitudes $e\phi/T_e \approx 0.15$ (see Fig. 13). It then follows that

$$p_\perp \lambda_D \approx \sqrt{\left|\Delta\omega_{\text{TP}}/2\omega_p\right|}. \tag{49}$$

Besides the amplitude threshold condition, as shown in Fig. 17, the mode must[46] fit within the LW's width,

$$p_\perp > k_0/(2F). \tag{50}$$

Eq. (9) implies that $\Delta\omega_{\text{TP}}/\omega_b \approx 0.2$ (see Fig. 5), while $e\phi/T_e \approx 0.15$ implies that

$\omega_b/\omega_p \approx 0.1$. Therefore $\Delta\omega_{\text{TP}}/\omega_p \approx 0.02$, and Eq. (49) implies $p_\perp \lambda_D \approx 0.1$. Since

$k_0\lambda_D \approx (v_e/c)\sqrt{n_c/n_e} \approx 0.17$, Eq. (50) is well satisfied for $F = 4.5$.

Because $\Delta\omega_{\text{TP}} \sim \sqrt{\phi}$ is a weak nonlinearity, solutions of Eq. (24) are not expected to attain scales much smaller than determined by the spectral width of unstable modes, as given by Eq. (49). This implies that the angular width of TPMI excited Langmuir waves, $\Delta\theta$, scales as $(1/k_0\lambda_D)\sqrt{\Delta\omega_{\text{TP}}/\omega_p}$, and for the above parameters, $\Delta\theta \approx 1/2$.

Since backscattered light that couples to obliquely propagating LWs will rapidly diffract out of the laser beam, most TPMI daughter LWs will not be amplified by the basic SRS process. Instead they play the role of an energy sink and hence a saturation mechanism for SRS.

## IV. DISCUSSION

Three simple estimates were presented, as to the importance of the Langmuir wave (LW) trapped particle frequency shift, $\Delta\omega_{TP} < 0$, for stability. The first compares $\Delta\omega_{TP}$ with the ponderomotive force induced shift (Fig. 6). This only involves the wave's normalized amplitude, $e\phi/T_e$ and wavenumber, $k\lambda_D$. The second compares the trapped particle modulational instability (TPMI) growth rate, $\gamma_{TPMI}$, with that of the Langmuir wave ion acoustic decay instability, LDI, (Fig. 15). This requires another parameter, $c_s/v_e$, the ratio of ion acoustic to electron thermal speeds. The third compares the TPMI and LDI amplitude thresholds (Fig. 22), which in addition depends upon the normalized rate of escape of trapped electrons, $v/\omega_p$, and the ion acoustic damping coefficient, $v_{ia}$. These comparisons all suggest regime change at $k\lambda_D \approx 0.3$ due to the rapid increase of $\left|\Delta\omega_{TP}\right|$ (Fig. 5) with increasing $k\lambda_D$. There is an intermediate wavenumber regime, *e.g.*, for small amplitude waves $0.33 < k\lambda_D < 0.46$, in which TPMI cannot be three-dimensional since fluctuations purely along the wave direction (*z*) have negative dispersion and hence are stable. The nonlinear saturation of TPMI will then consist of elongated structures in the *z* direction. The *x-y* extent is estimated as small compared to the experimental[9] beam width. One-dimensional models are TPMI stable in this regime.

Nonlinear dispersion, another aspect of $\Delta \omega_{TP}$, becomes significant as $\nu/\omega_p$ and the threshold value of $e\phi/T_e$ increase: dispersion decreases, until TPMI is not possible for $\nu/\omega_p > 0.005$. The positive dispersion amplitude boundary is lower than the loss of resonance boundary (Fig. 15), and thus determines the TPMI damping limit (Fig. 22).

Assumptions have been made concerning slow variation so that the modulational model, Eq. (14), is valid. Slow variation in time requires that the TPMI growth time is large compared to a bounce period, $2\pi/\Delta\omega_b$. This is easily satisfied since even at $k\lambda_D = 0.45$, $|\Delta\omega_{TP}|$ is only a fraction of $\omega_b$, and at most, $\gamma_{TPMI} = |\Delta\omega_{TP}|/4$. Slow space variation requires that the wavevector of the most unstable fluctuation, **p**, is small enough so that the diffractive approximation, Eq. (20), to full dispersion, Eq. (25), is accurate. Physically, this may be too small to be useful when $k\lambda_D \approx 0.33$, the point where the *z* dispersion coefficient changes sign for small values of $e\phi/T_e$. On the other hand, TPMI relevant values of $e\phi/T_e$ may lower this *z* diffractive transition point (Fig. 12) far enough below wavenumbers of interest so that the diffractive model is valid, albeit with amplitude dependent coefficients (Figs. 12 and 13).

The modulational model may be challenged by rapid spatial variation due to the nonlinear LW response itself. For example, even if the source, $\phi_0$, varies slowly in space, there is a jump in nonlinear response at the response threshold, as shown in Fig.

16. In space, there will be a transition from the large amplitude resonant response regime, with weak Landau damping and large $\omega_b$, to a small amplitude, non-resonant response regime, with strong Landau damping and small $\omega_b$. The assumed local relation between $\Xi$ and $\phi$, Eqs. (4) and (5), will break if $\omega_b < \mathbf{v}_p \left| \partial/\partial z \ln|\phi| \right|$, with $\mathbf{v}_p$ the LW phase velocity. A simple transport model for $\Xi$, with advection velocity $\mathbf{v}_p$, and relaxation to Eqs. (4) and (5) at rate $\omega_b$, has revealed qualitative changes in behavior when applied to backward SRS modeling due to this effect[47]. Note, however, that even in the strongly trapped and $\omega_b > \mathbf{v}_p \left| \partial/\partial z \ln|\phi| \right|$ regime, the response of $\phi$ as given by Eq. (15) (supplemented by source and nonlinear damping as in Eq. (28)) may be quite different from the local equilibrium solution, Eq. (13), and in this sense non-adiabatic. All that is required is that $\phi_0$'s time variation is fast compared to $\Delta\omega_{\mathrm{TP}}$. Since there is typically a large gap between $\Delta\omega_{\mathrm{TP}}$ and $\omega_b$ (see Fig. 5), the LW response may simultaneously be non-adiabatic and still accurately given by the modulational model.

Comparison of LDI and TPMI thresholds requires a re-evaluation of the LDI threshold due to the primary LW's trapped particle frequency shift. If this effect were ignored, then LDI (with $\nu_{ia} = 0.1$) would require a non-resonant primary LW with $e\phi/T_e = O(1)$ when $k\lambda_D > 0.36$. Instead, the LDI threshold may be dramatically lowered (Fig. 21), making for more interesting competition with TPMI.

Of course, one need not invoke competition with TPMI to explain the experimentally observed[9] regime change, whose principle qualitative feature is the change from a coherent, LDI-like LW spectrum, to an incoherent spectrum, at $k\lambda_D \approx 0.32$. For example, as an alternative mechanism, the SRS daughter Langmuir wave may simply fall below the LDI threshold as $k\lambda_D$ increases.

Since $\nu \propto 1/(\text{beam diameter})$, and since the LDI/TPMI threshold crossover is approximately controlled by $\left(\nu/\omega_p\right)^4 / \nu_{ia}$, a TPMI significant reduction of $\nu$ is achievable by doubling the optic f/#. For example, if $\nu/\omega_p$ were reduced from 0.004 to 0.002 (see Fig. 22), the LDI/TPMI crossover is lowered to $k\lambda_D \approx 0.30$ from $k\lambda_D \approx 0.33$. If such a decrease in $k\lambda_D$ at loss of LDI were seen experimentally, this would be one vote in favor of competition with TPMI as the operative mechanism since the value of $\nu$ has no effect on the linear LDI threshold.

Quantitative estimates of $\nu$ may remain elusive, as it pertains to SRS modeling, since its value must depend on details of the daughter Langmuir wave geometry, and that geometry is quite variable. For example, if the SRS spatial gain rate is large enough, the LW width can be smaller than that of the laser beam due to gain localization. Even if this is not the case, in a spatially incoherent laser beam the regions of large intensity fluctuations (speckles) where SRS thrives, are only approximately diffraction limited, and other processes such as forward stimulated Brillouin scatter may change the speckle length (and width) with beam propagation[48], implying significant uncertainty in the basic LW geometry. However, for given LW geometry, the value of $\nu$ has significance beyond that as a parameter in a 1D modified Vlasov equation: if the extrapolation of the LW damping rate to small values of $k \mathrm{v}_e / \omega_b$ (at fixed $k$) implies a finite residual damping, then it is that residual which determines an effective value for $\nu$ and a lower bound to the TPMI threshold. In the context of the current 1D model, that residual is determined by the free parameter $\nu$, and the function $\Delta$, as in Eq. (5).

**Acknowledgements**

This work is supported by the Department of Energy, under contract W-7405-ENG-36. Discussions with J. Kline and D. Montgomery, concerning interpretation of their experimental data, are greatly appreciated.

---

weakly trapped regime, a LW wavepacket with transverse length scale $L >> 1/k$ has a damping greater than its 1D evaluation through an entirely different mechanism: its Landau damping is not evaluated at k, but at a slightly larger wavenumber,

$$\approx k + \pi k \Big/ (kL)^2 .$$

[18] See Fig. 1 of Ref. 10. The particular class of related BGK modes is also discussed there.

[19] J. P. Holloway and J. J. Dorning, Phys. Lett. A **138**, 279 (1989); Phys. Rev. A **44**, 3856 (1991).

[20] It is assumed that $v/v_e > 1$, so that $f_0''(v/v_e) > 0$, as is the case for the Langmuir and electron-acoustic modes.

[21] T. P. Coffey, Phys. Fluids **14**, 1402 (1971).

[22] W. L. Kruer, in The Physics of Laser Plasma Interactions, 1st edition, edited by D. Pines ~Addison-Wesley, New York, 1988, Chap. 9, p.104.

[23] See section III.B.4 of Ref. 10.

[24] Because of ion inertia, the actual ponderomotive induced shift tends to be smaller.

[25] This boundary is lowered if there is a broad spectrum of Langmuir waves such that an otherwise trapped electron is lost by diffusion in a time $< 1/\omega_b$. The analysis presented here is for the case of a coherent wave.

[26] The expression for Re $\Xi$, however, as given by Eq. (4), may be valid for $\phi$ larger than its loss of resonance value for a particular value of $k$, because the definition and evaluation of $\Xi$ does not require a plasma resonance.

[27] The apex of this region is not quantitatively accurate since Eq. (9), and hence Eq. (12), breaks down near loss of resonance.

[28] Mathematically, all that has been proven is that for $k\lambda_D > 0.53$, there do not exist finite amplitude traveling wave solutions which bifurcate from thermal equilibrium. See Ref. 11.

[29] The anharmonic components of the exact solution can be quite small. See Eq. (60) in Ref. 10.

[30] The external potential and relaxation to the background distribution function are omitted here.

[31] Ref. (7) provides the estimate that the temporal variation is "slow" if its associated frequency is small compared to the bounce frequency.

[32] B. I. Cohen and A. N. Kaufman, Phys. Fluids **21**, 404 (1978). There is an enormous literature on the derivation of nonlinear wave equations, going back at least to the seminal work of G. B. Whitham, Proc. Roy. Soc (London **A283**, 238 (1965)), which uses a Lagrangian based, variational approach. The point of view taken by Cohen and Kaufman, and this work, (dissipation and a source are explicitly allowed in section III.C.1), is that the wave is supported by a finite amplitude source, which is balanced by dissipation. The equilibrium , harmonic approximation, response to a monochromatic source is determined by the nonlinear dielectric function, $\varepsilon$, and thus its Taylor series naturally describes slow modulations about such an equilibrium.

[33] I. B. Bernstein, J. M. Greene, and M. D. Kruskal, Phys. Rev. **108**, 546 (1957).

[34] L. D. Landau, J. Phys. (*U.S.S.R.*), **10**, 25 (1946).

[35] It may be shown that this definition, and those for dispersion coefficients, are identical to the conventional definitions in terms of derivatives of $\omega$ with respect to $k$, when these variables are related via the linear dispersion relation, $\mathrm{Re}\,\varepsilon_0(k,\omega) = 0$.

[36] Terms which are proportional to products of $\partial/\partial t$ and space derivatives are ignored.

[37] So long as there is positive dispersion.

[38] Dwight R. Nicholson, in *Introduction to Plasma Theory*, **1**st edition, edited by Sanborn C. Brown (John Wiley & Sons, New York, 1983), Chap. 7, p.183, equation (7.352), with the parameter $\eta$ set to unity. This parameter may be defined by the magnitude of the normalized low frequency electron density fluctuation, $\delta n_{el}/n_e$, induced by the ponderomotive force: $-\delta n_{el}/n_e = (0.25/\eta)(k\phi)^2$, in thermal units.

[39] *Locus citatus*, page 182, equation (7.347).

[40] Any direction for the LDI daughter wave, in principle, can support the instability, though absent strong damping anisotropy, backscatter is favored over sidescatter, both because the intrinsic growth rate is a maximum in the former and because of the highly elongated geometry of the laser beam used in experiment[9]. While this needs to be re-examined, in view of anisotropic trapped particle effects, since experimental data[9] is in quantitative agreement with LDI backscatter, this is the point of view adopted here.

[41] These considerations strictly apply near threshold. If the LDI daughter wave attains finite amplitude, its damping may decrease.

[42] For example, see Fig. 9 of Ref. 10.

[43] As given, *e.g.*, by equation (15) in H. A. Rose, Phys. Plasmas **10**, 1468 (2003).

[44] If the average laser intensity, $\langle I \rangle$, is above its critical value for SRS, then the power gain over a speckle length is at least as large as the ratio of a particular speckle's intensity to $\langle I \rangle$ (see H. A. Rose and D. F. DuBois, Phys. Rev Lett. **72,** 2883 (1994)). Since the most probable speckle intensity (absent self-focusing) is about $3\langle I \rangle$ (see J. Garnier, Phys. Plasmas **6**, 1601 (1999)), this condition is then easily satisfied.

[45] J. Kline, private comm., (2004).

[46] The spectrum of unstable modes cuts off at a wavenumber only $\sqrt{2}$ larger.

[47] H. A. Rose, BAPS, 2001, QP1.145

[48] P. M. Lushnikov and H. A. Rose, "Instability versus equilibrium propagation of laser beam in plasma", accepted for publication by Phys. Rev. Lett., April 2004 [e-print xxx.lanl.gov/physics/0312055].